\documentclass[12pt]{article}

\usepackage{graphicx}
\usepackage{subfigure}
\newfont{\feff}{cmti10}
\begin{document}

\title{Large-Scale Coherence and Law of Decay of  
Two-Dimensional Turbulence}
\author{ Victor Yakhot, John Wanderer\\
Department of Aerospace and Mechanical Engineering,\\ Boston
University, Boston 02215}

\maketitle \begin{abstract}
\noindent   
At the short times,
 the enstrophy $\Omega$ of a two-dimensional flow, generated by a random 
Gaussian initial condition 
decays as $\Omega(t)\propto t^{-\gamma}$  with $\gamma\approx 0.7$. After that,
the flow undergoes transition to a different state  
characterized by the magnitude of  the decay exponent $\gamma\approx 0.4$ (Yakhot, Wanderer, Phys.Rev.Lett.{\bf 93}, 154502 ~(2004)). 
It is shown that the very existence of this  transition and 
various characteristics of evolving flow crucially depend upon 
phase correlation between the large- scale modes  containing only a 
few percent of  total enstrophy.

\end{abstract}
\newpage

\noindent \noindent  Two-dimensional (2D) turbulence is a simplified 
 system mimicking such important  phenomena, 
 as ubiquitous vortex generation,   
 in atmospheric flows.  The vortex formation as a dominant process  of 2D hydrodynamics  has been recognized by Onsager more than half a century ago$^{1}$ 
and, thanks to  obsession of   
TV stations with the  weather forcast, the relevance of spontanious emergence of 
strong vortices out of quiescent background for the world we live in,  can be verified at any time of the day. 
Many features of the vortex formation in decaying 2D 
turbulence have been undestood due to numerical and theoretical works of McWilliams and collaborators$^{2-5}$, 
Benzi et. al.$^{6}$ and many others.  Still, many new features have 
recently been discovered 
and some important questions remained unanswered.

The problem of decay of two-dimensional (2D) turbulence considered in our recent paper$^{7}$  and 
the one, we are interested in here,  is formulated in a following way.
Consider a time evolution of an initial velocity field, $u(x,y,0)={\bf u}_{0}$ 
defined on a two-dimensional square such that $-L\leq x,y\leq L$. 
The field $u_{0}$ is a Gaussian random noise supported  in the Fourier space in the vicinity 
of $k=k_{0}(t=0)=O(\frac{N\pi}{L})$ with $\overline{{\bf u}^{2}_{0}}=O(1)$ and $N=const=O(1)$.
At the initial instant, $t=0$, the enstrophy is $\Omega=\overline{\omega^{2}}\equiv \omega_{0}^{2}=\overline{(\nabla \times {\bf u})
^{2} }=O(1)$. The square is large meaning that we are interested in the limit $\nu\rightarrow 0$ and 
$k_{0}(t=0)L\rightarrow\infty$. Still, the box is finite so we will be able to
study both short time,
when $k_{0}(t)L\gg 1$, and  
long time asymptotics when $k_{0}(t)L\approx 1$. In all our simulations the initial kinetic energy of the flow
$K(t=0)=\frac{1}{2}\rho \int v^{2}({\bf x},t=0)d{\bf x}=\frac{1}{2}$ and the fluid density $\rho=1$.
This set up with the basically structureless initial field is ideal for 
investigation of the details of the structure emergence in the course of the flow evolution.

According to  a recent theory$^{8}$,  in an infinite system  ($L\rightarrow \infty$), 
the universal asymptotic law of
enstrophy decay is $\Omega\propto t^{-\gamma}$ with $\gamma={\frac{2}{3}}$ 
which is close to the short -time results of simulation by Chasnov$^{9}$ ($\gamma\approx 0.7-0.8$) and our own$^{7}$
$\gamma\approx 0.7$.  At the later times, after the large-scale (small-wave-vector) modes are somewhat populated,
the system undergoes transition to a gas or liquid of well separated vortices and, simultaniously, the magnitude 
of the exponent $\gamma$ crosses  over$^{7}$  to $\gamma\approx 0.4$,  well-known from studies of evolution of an initially created ensemble of point vortices$^{1-6}$. The theory developed in Ref.[7] gave for the number of  vortices in the system 
$N(t)\propto t^{-\xi}$ with $\xi=4/5$ and $\gamma=2/5$ very close to numerical data. This theory  was based  on the work by Carnevale et al$^{3}$  which
led to the relation  $\gamma=\xi/2$ with an undetermined magnitude 
of exponent  $\xi$. 
Although a tentative 
correlation between the cross-over to $\gamma\approx 0.4$ with population of the 
large-scale modes was mentioned in Ref.[7], 
the details of the process remained obscure. 
All theories treated the vortex merger process locally  as a binary ``chemical reaction'' 
$2n\rightarrow n$ with the constant peak vorticity before and after collision. 
The role of the large-scale patterns on  scales $l=O(L)$  in this process was completely ignored.
Below we report a surprising 
discovery: the very existence of  
transition to the long-time asymptotic  ($\gamma=0.4$) of the enstrophy decay 
and the energy  of evolving vortices crucially depend
upon phase correlation (coherence) of the large-scale modes containing only a tiny fraction of the total 
enstrophy $\Omega$.
 
The Navier-Stokes equations with the $O(\nu_4 \nabla^{4}{\bf v})$ hyper-viscous dissipation  terms were simulated 
using a pseudo-spectral method. The initial random field was Gaussian with energy spectrum 
$E(k)=a_s \frac{K}{\rho  k_p} \left(\frac{k}{k_p}\right)^{2 s +1} e^{-\left( s+ \frac{1}{2}\right)\left(\frac{k}{k_p} \right)^2 } $ 
where $a_s=\frac{(2s+1)^{(s+1)}}{2^s s!}$ and $s=3$ as in Ref.[9].
  The parameters of the simulations are given in a table of Ref.[7]. 

%\begin{table}[h]
%\center
%\begin{tabular}{|c|c|c|c|c|c|}
%\hline
% N & time-step & $\frac{2K}{\rho}$ & $\Omega(t=0)$ & $k_p$ & $\nu_4$ \\
%\hline
%\hline
% $512$ &$ 0.001$ &$ 1$ &$ 293$ &$ 16$ & $21.74x10^{-9} $ \\
%\hline
% $1024$ &$ 0.0005$ &$ 1$ &$ 1170$ &$ 32$ & $2.1x10^{-10} $ \\
%\hline
% $2048$ & $0.00025$ &$ 1$ &$ 1170$ &$ 32$ & $9.6x10^{-11} $ \\
%\hline
% \end{tabular}
% \caption{Numerical simulation p arameters\label{fig:num_params}}
%\end{table}

On Fig. 1  the time 
evolution of normalized enstrophy $\Omega(t)/\Omega(t=0)$ is  presented. The curve A  is the one 
obtained from the $512^{3}$ simulations reported in Ref.[7]. The cross-over from 
from the short-time magnitude of the decay exponent $\gamma\approx 0.7$ 
to the long -time asymptotics $\gamma\approx 0.4$   happening at the dimensionless 
time $T=t\omega_{0}\approx 120$ is clearly seen (curve A). To investigate the role of the large-scale 
modes in this phenomenon we, at the transition time $T\approx 120$, 
randomized the phases of the first few modes ($k\leq 5$ ) (curve R)  and let the flow evolve. 
It is important to stress that this procedure   influenced neither total enstrophy nor energy in the system.
Still, the observed response of the flow to this perturbation  was 
quite dramatic: 
\begin{figure}[h]
  \center
\subfigure[Time evolution of total  enstrophy $\Omega$. A.~ Unperturbed field. R.~ effect of randomization of the modes with $k\leq 5$;  N.~$\Omega(t)$ at $T\geq 120$ with $v(k)\rightarrow -v(k)$ ($k\leq 5$); RR.~evolution after second randomization at $T\approx 470$.]{\includegraphics[height=9cm]{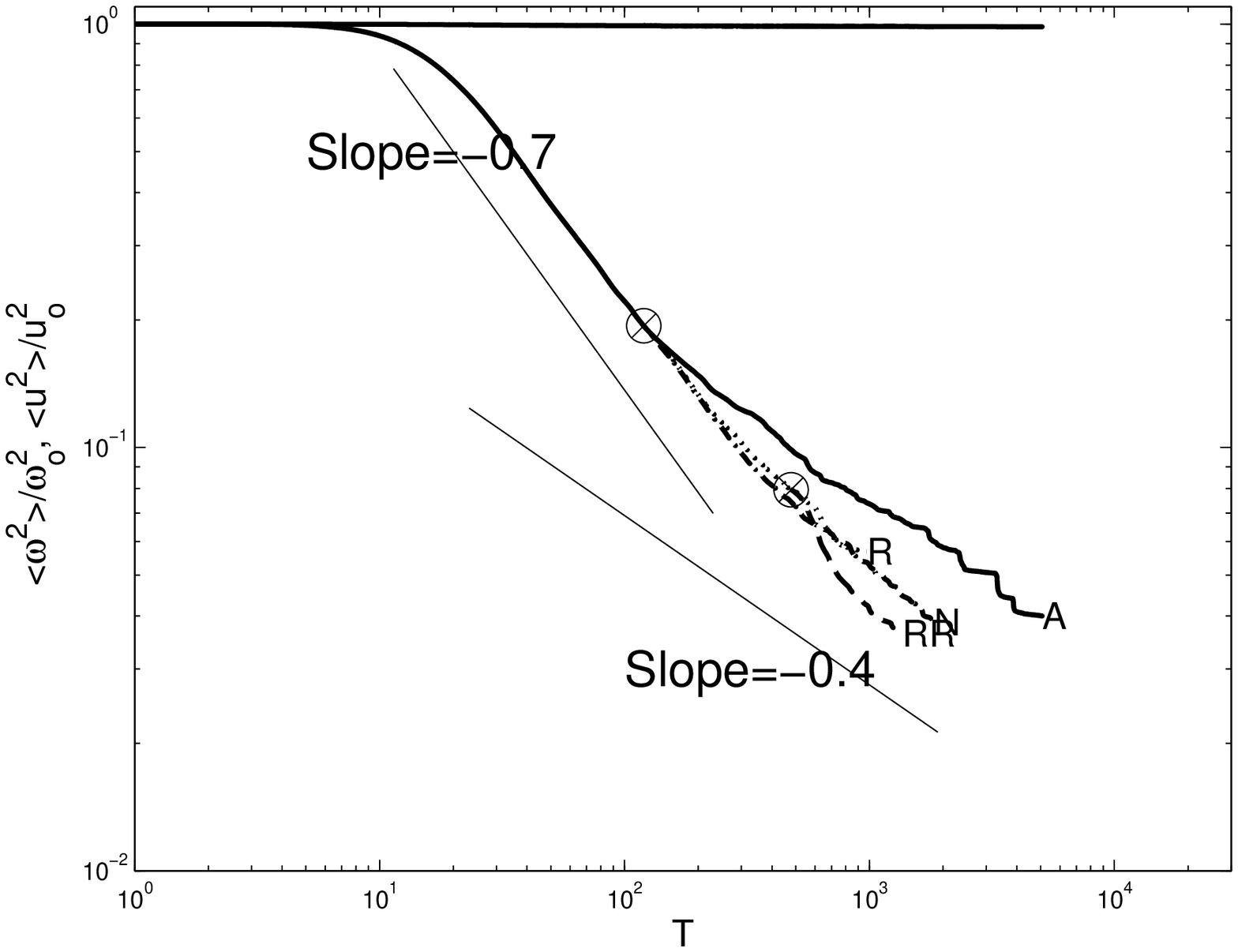}}
   \subfigure[A. The same as on 1a;~~R4: Evolution after randomization of the modes with $k\leq 4$.]{\includegraphics[height=9cm]{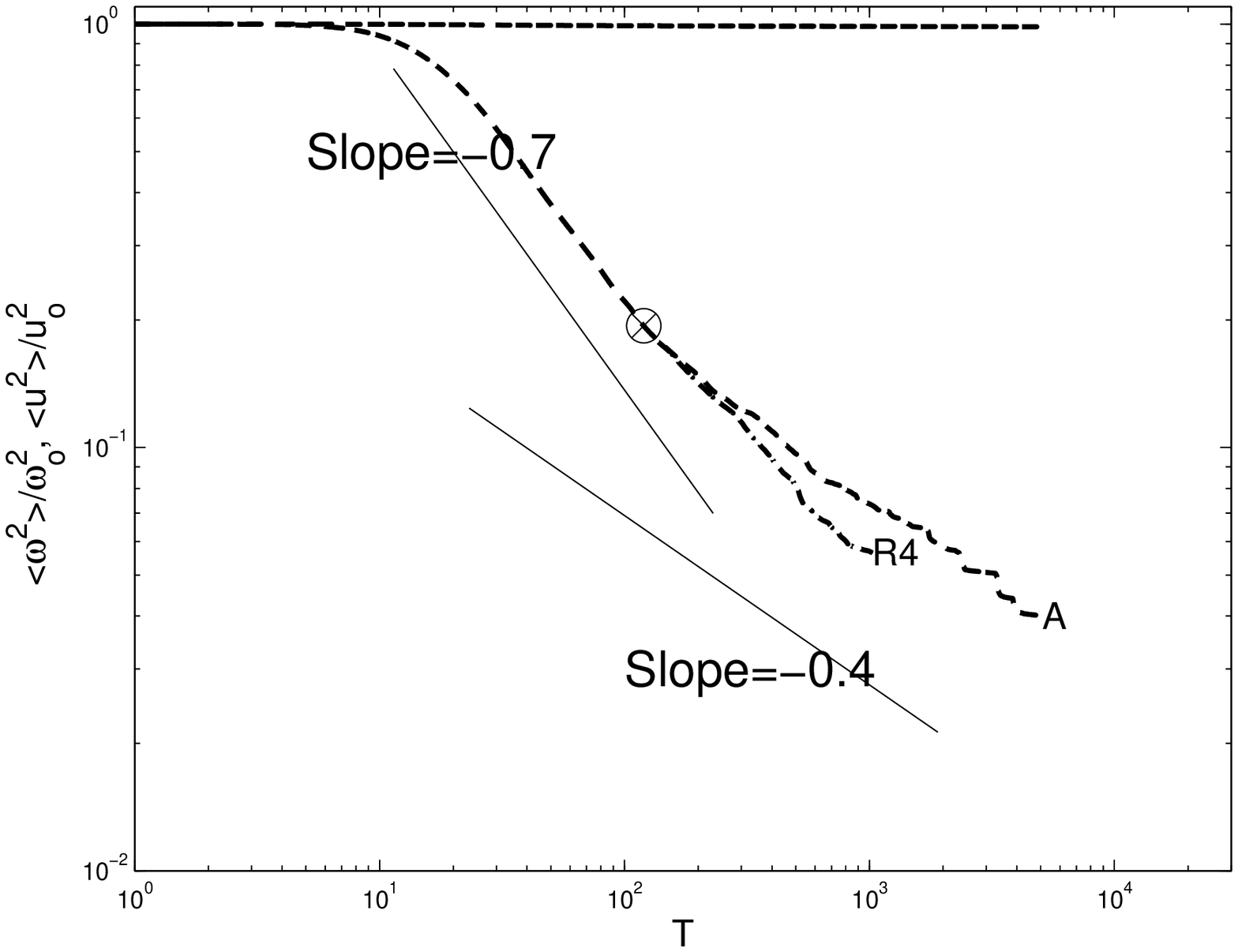}}
  \caption{Enstrophy vs Time}.
  \label{fig:omega_decay}
\end{figure} the randomization of  phases of the modes, containing only a few percent of the total enstrophy,  prevented 
transition from happening. 
 Then, 
following the discussion with J.McWilliams, instead of randomizing, we at the same instant  $T=120$
inverted the signs of the same modes in the unperturbed run $A$, i.e. made a transformation 
$v_{i}(k)\rightarrow -v_{i}(k)$ with $k\leq 5$. The  effect of this procedure 
on the enstrophy evolution can be seen on Fig. 1a (curve N):  the curves $R$ and $N$ are almost indistinguishable.  After some time  ($T\approx 470$ ) the perturbed flow ($R$)  developed the  large-scale coherent motions and a tendency to crossover to the long-time asymptotics  ($\gamma\approx 0.4$) could be detected. 
At this instant , again,  the phases of the modes with $k\leq 5$ have been randomized.  The result of this perturbation
is shown on the curve $RR$ of Fig. 1a.  We can see that the  second randomization too,  forced  the enstrophy  to decay  much  faster than 
$\Omega\propto t^{-0.4}$ observed  in the dynamically unperturbed run $A$.
At this point we can make a definitive statement:
the  phase correlation ,  dynamically established as a result of the large-scale evolution,   is crucial for the very existence of  transition to the long-time enstrophy decay regime. 

If the previous experiment established strong coupling between  large ($k \leq 5$)  
and small ( enstrophy containing) scales , then the next  natural question to ask is:  what is the minimal length-scale of  the energy-containig structures influencing such small-scale phenomenon as  the enstrophy decay process?  To answer this question, we broadened our search. The phase randomization of the modes $v(k)$ with $k\leq 2$ did not show any effect on the decay process.  
The result 
of randomizing the phases of the modes with $k\leq 4$ containing at $T\approx 120$ less than five percent of total enstrophy $\Omega$, is shown on Fig. 1b: the transition to the long -time behavior is delayed.  

The time-dependence of the fourth and sixth-order moments are shown on Fig. 2.  As we see,  the initially gaussian random flow ($t=0$) develops strongly non-gaussian tails characteristic of the small-scale vortex formation also observed in Ref.[10].  It is interesting that  at the long times,  the normalized  high-order moments of the unperturbed vorticity  field ($A$) approach  a close to steady state.  Thus, in  in accord with Ref.[11],  we can write an exact  equation for probability  density $P(X)\equiv P(\frac{\omega}{\omega_{rms}})$  in terms of  conditional expectation value of  vorticity dissipation rate $Q(X)$ for a given magnitude  of $X$.  A simple model  $Q(X)\propto (1+kX^{2})$ proposed and numerically verified in Refs. [11]  gives:

\begin{equation}
P(X)=\frac{C}{(1+\kappa X^{2})^{1+\frac{1}{2\kappa}}}
\end{equation}

\noindent  where $C$ is a normalization constant.   After the transition ( $T>120$),  the maximum value of vorticity in the evolving unperturbed field $\omega_{max}/\omega_{0}\approx 60\approx const$ and the expression (1)  is valid in the interval $X_{max}\leq \omega_{max}/\omega_{rms}(t)$. For $X>X_{max}$, the probability density is cut off, i.e. we can set  $P(X)\approx 0$.  It is due to the cut-off $X_{max}$, the probability density (1) is consistent with the finite moments of vorticity field. 
Since $\omega_{rms}=O(t^{-0.2})$, the magnitude of $X_{max}\propto t^{0.2}$ does not appreciably varies during the time of our  simulation and  can be set $X_{max}\approx const$.  The data (curves $A$) reported on Fig. 2 are reasonably well fitted by $\kappa\approx 1$ and $X_{max}\approx 30$.  As $\kappa\rightarrow 0$, the relation (1) approaches  a Gaussian and, since the evolution of the normalized moments is slow,
the curves $A$ of Fig.2 can be represented by (1) with the time-dependent parameter $\kappa(t)$. 
As we see from Fig.2, during the time of simulation, the sixth-order   moments of  perturbed runs $R$ and $N$ do not reach  steady state meaning that the strong vortices decay with the exponent $\gamma_{s}<0.7$.

To make the  discussion easier we will be dealing with three fields: 1. unperturbed (A); 2. randomized (R); and 3.
sign-reversed (N). The time-evolution of vorticity field in all three runs (A;~R~N) 
is shown on Figs. 3-5. On Figs.3(A;R;N;)  the  vorticity fields at the transition time 
$T\approx 120$ right after randomization (Fig.3R) and sign reversal 
(Fig. 3N) are compared with the original unperturbed  field (Fig. 3A).  In all three plates we can  identify the
 small-scale strong vortices which  have not been affected by the randomization or  sign-reversal. 
The patterns  between these structures  are quite different, though. The time-evolution of these fields presented on 
Figs.4-5 for the times $T\approx 480$ and $T\approx 975$, respectively, 
leads to substantially  different patterns: the larger  number of strong vortices (also see below) 
can be seen in the field $A$, while the  evolution of the fields $R$ and $N$ results in the smaller number 
of vorticies. While during this evolution, the maximum value of vorticity stayed more or less unchanged 
($\omega_{max}\approx 65-60$), the number of strongest vortices having the peak value of vorticity 
$\omega_{o}\approx 0.8\omega_{max}$ (not shown here due to lack of space) 
in the unperturbed field $A$ was much larger (factor 1.5-4, depending on time $T$) 
than that in the corresponding perturbed runs $R$ and $N$.

In Fig. 6
we show the time-evolution of maximum velocity $v_{max}$ in all three cases as a function of time. 
We can see that after some transient,  the velocity in the unperturbed field $A$ reaches the values  $v_{max}\approx 3.5-4.$,
while in the time interval $120<T<500$ the velocity in  perturbed fields $R$ and $N$ is smaller:
$v_{max}\approx 2.7-3.$. 
It is interesting that by the time $T\approx 600$, the maximum velocity of the field $R$ is more or less recovered
while that in  a stronger perturbed  field $N$  remained substantially smaller than that in the $A$-field during the remaining time of the simulation. 

\noindent  The strong correlation between  small and large-scale structures in the small-scale forced 
2D  turbulence was discovered  in Refs. [12]-[14]. Two different flow regimes were identified: at the short times, when the time-dependent integral scale $l_{I}\ll L$,  the inverse cascade led to generation of  two -dimensional turbulence characterized by the Kolmogorov spectrum  and close to gaussian magnitudes of the even-order moments velocity differences.   The most interesting effect happened at the later times when the modes with $k\approx 1/L$ became populated:  very strong vortices,  were simultaneously formed at the forcing scale $l_{f}\approx 1/k_{f}$.  It was demonstrated$^{13}$ that in the case 
of forced turbulence,   the vortices were created  at the centers of the large-scale, slowly -varying  patterns where velocity $v\approx 0$.  Thus, the Bose condensate served as an almost steady 
well-organized matrix  facilitating the vortex merger process. The possible generality of this phenomenon was demonstrated by the experiments of Paret and Tabeling$^{14}$ in their 
 investigation of   time-evolution of the  two-dimensional flow of mercury generated by magnetic field. 
 
 It is important to stress that, unlike in  the forced flow considered in Refs. [12]-[14] where  vorticity $\omega_{max}$ 
 of the strongest vortices continuously grew with time, in decaying turbulence studied here, 
   $\omega_{max}(t)\approx const \approx 60$. If this feature persists for a very long time,
 eventually  only two strong vorticis  of the radius  
 $a_{\infty}\approx [K/(\rho\omega_{max}^{2})]^{\frac{1}{4}}$
 will be  left in agreement with Ref.[1]. For the  flow considered in this paper 
 $a_{\infty}\approx 0.1$. Thus, the final vortex  occupies a tiny fraction ($\approx 10^{-3}-10^{-4}$) of the flow.  
 The flow velocity in this configuration will reach $v \approx a_{\infty}\omega_{max}
 \approx \sqrt{\omega_{max}}[K/\rho]^{\frac{1}{4}}\approx  5-10$ which is much larger than that in the initial random state.   Thus, the 2D turbulence decay  leads to formation of localized   high-energy structures out of the initially quiescent background.
 
 To conclude: the main result of the present study is:  we have showed that the 
cross-over  between the two asymptotics of  enstrophy decay in 2D turbulence strongly depends 
upon fine details of the large-scale flow features  containing
  very small fraction of total enstrophy $\Omega$. The dynamic cause of this transition is not yet clear.  
However, we can conclude that  fine, dynamically evolved,
phase-correlation of  the large-scale modes  leads to a slower 
enstrophy decay and  formation of more energetic flow  structures. This points to  a strong coupling 
between  the small-scale and large-scale (condensates) structures  in 2D turbulence.  At the present time, we do not have a theory leading to this coupling and do not fully appreciate  it's consequences. Based on previous examples of coherent-random states interaction  (superfluidity, superconductivity and many others) ,
the theory of the effect observed in this paper may be a substantial theoretical challenge.
The possibility of this coupling in forced 2D turbulence was discussed by Polyakov$^{15}$ in his conformal theory of two-dimensional turbulence.  Since the geometry-dependent large-scale patterns cannot be universal, the  question of 
 universality of the transition  is extremely interesting.  The possible  role of the inter-scale
 interaction, observed in this work,    may be of great interest for meteorological applications. 

\noindent We are grateful to J. McWilliams and A.Polyakov for valuable comments and suggestions.

\noindent{\bf References.}\\
\noindent 1.~L. Onsager, Nuovo Cimento {\bf 6}(2),279 (1949).\\
\noindent 2~. J.C. McWilliams, J.Fluid Mech {\bf 219}, 361 (1990).\\
\noindent 3.~G.F.Carnevale, J.C.McWilliams, Y.Pomeau, J.B.Weiss and W.R.Young, Phys.Rev.Lett.{\bf 66}, 2735 (1991).\\
\noindent 4.A.Bracco, J.C.McWilliams, G.Murante,A.Provenzale,J.B.Weiss, Phys.Fluids {\bf 12}, 2931 (2000).\\
\noindent 5.~M.V.Melander, J.C.McWilliams, and N.Zabusky, J.Fluid mech. {\bf 178}, 137 (1987).\\ 
\noindent 6.~R.Benzi, S.Patarnello and P.Santangelo, J.Phys.A {\bf 21}, 1221 (1988).\\
\noindent 7.~ V.Yakhot and J.Wanderer, Phys.Rev.Lett.,{\bf 93}, 154502 (2004).\\
\noindent 8.~V.Yakhot, Phys.Rev.Lett., {\bf 93}, 014502,  2004.\\
\noindent 9. ~J.R.Chasnov, Phys. Fluids {\bf 9}, 171 (1997).\\
\noindent 10. L.M. Smith V. Yakhot, Phys.Rev.E{\bf 55}, 5458 (1997).\\
\noindent  11. Ya.G. Sinai  V.Yakhot,  Phys.Rev.Lett., {\bf 63}, 1962  (1989);~V.Yakhot, S.Orszag, S. Balachandar, E. Jackson, Z.S.She, L.Sirovich, J.Sci.Comp.,{\bf 7}, 199 (1990).\\
\noindent 12.~ L.M. Smith and V.Yakhot, Phys.Rev.Lett. {\bf 71}, 352 (1993).\\
\noindent 13.~ L.M. Smith and V. Yakhot, J.Fluid Mech. {\bf 274} (1994).\\
\noindent 14~ J. Paret and P. Tabeling, Phys. Fluids {\bf 12}, 3126 (1998). \\
\noindent 15.~ A. Polyakov, Nucl.Phys. {\bf 386},  367 (1993).\\
%\noindent 13. R.H.Kraichnan and D. Montgomery, Reports Prog.Phys. {\bf 43}, 547 (1980)\\

%\noindent 17. A.S. Monin and A.M.Yaglom, vII, The MIT press, Cambridge, 1975\\
%\noindent 18. L.D. Landau and E. M. Lifshitz, Pergamon Press, Oxford, 1982\\
%\noindent 19. L.G. Loitsyanskii,Trudy Tsentr. Aero.-Gidrodyn. Inst., {\bf 3}, 33 (1939), (in Russian) \\

\begin{figure}[h]
 \center
 \includegraphics[height=9cm]{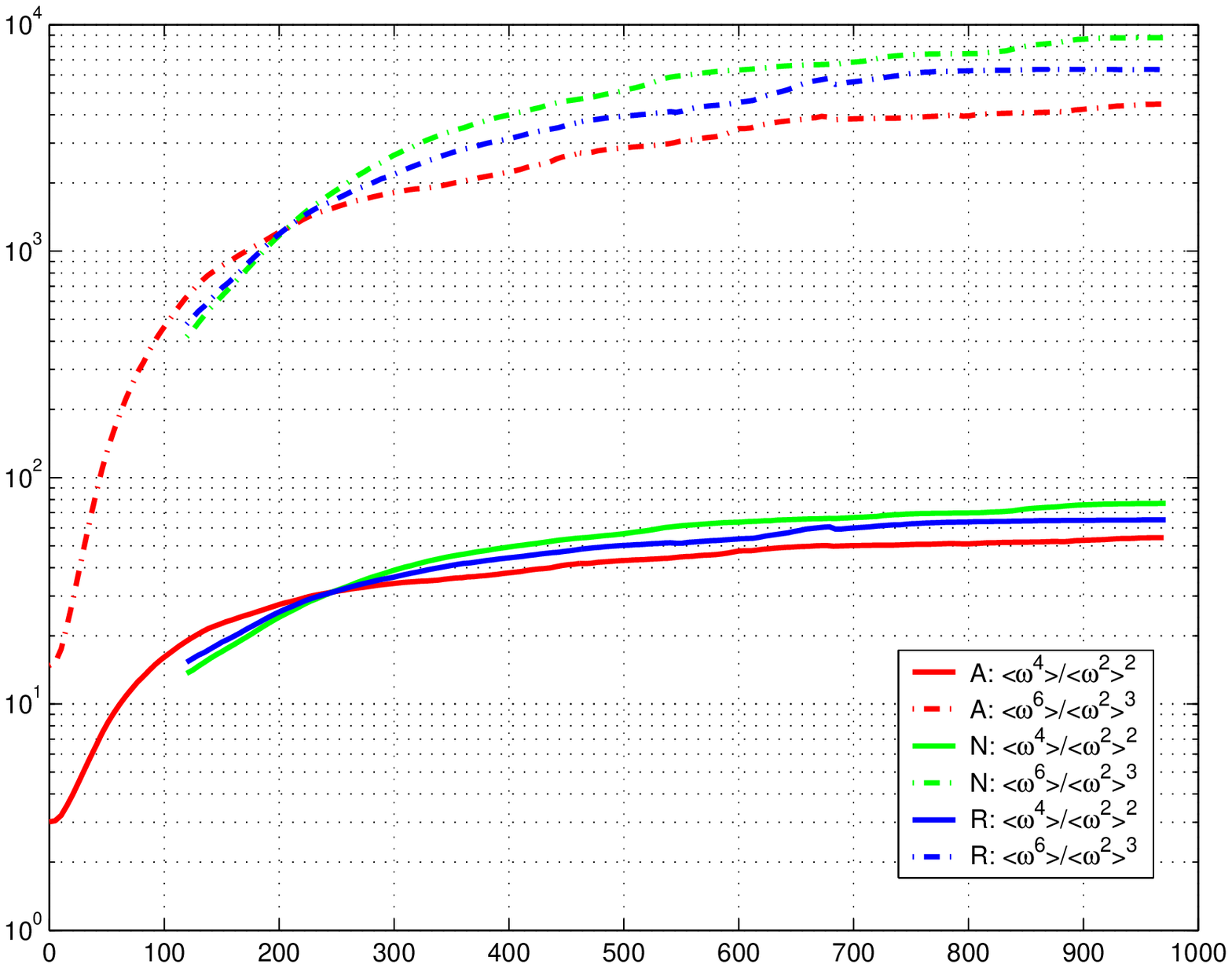}
  \caption{Time-evolution of vorticity moments for the  three runs }
  \label{fig:omega_moments}
\end{figure}
%\end{document}

\begin{figure}[h]
  \center
  \subfigure[A;~$\omega(x,y)$;~T=120. ]{\includegraphics[height=4cm]{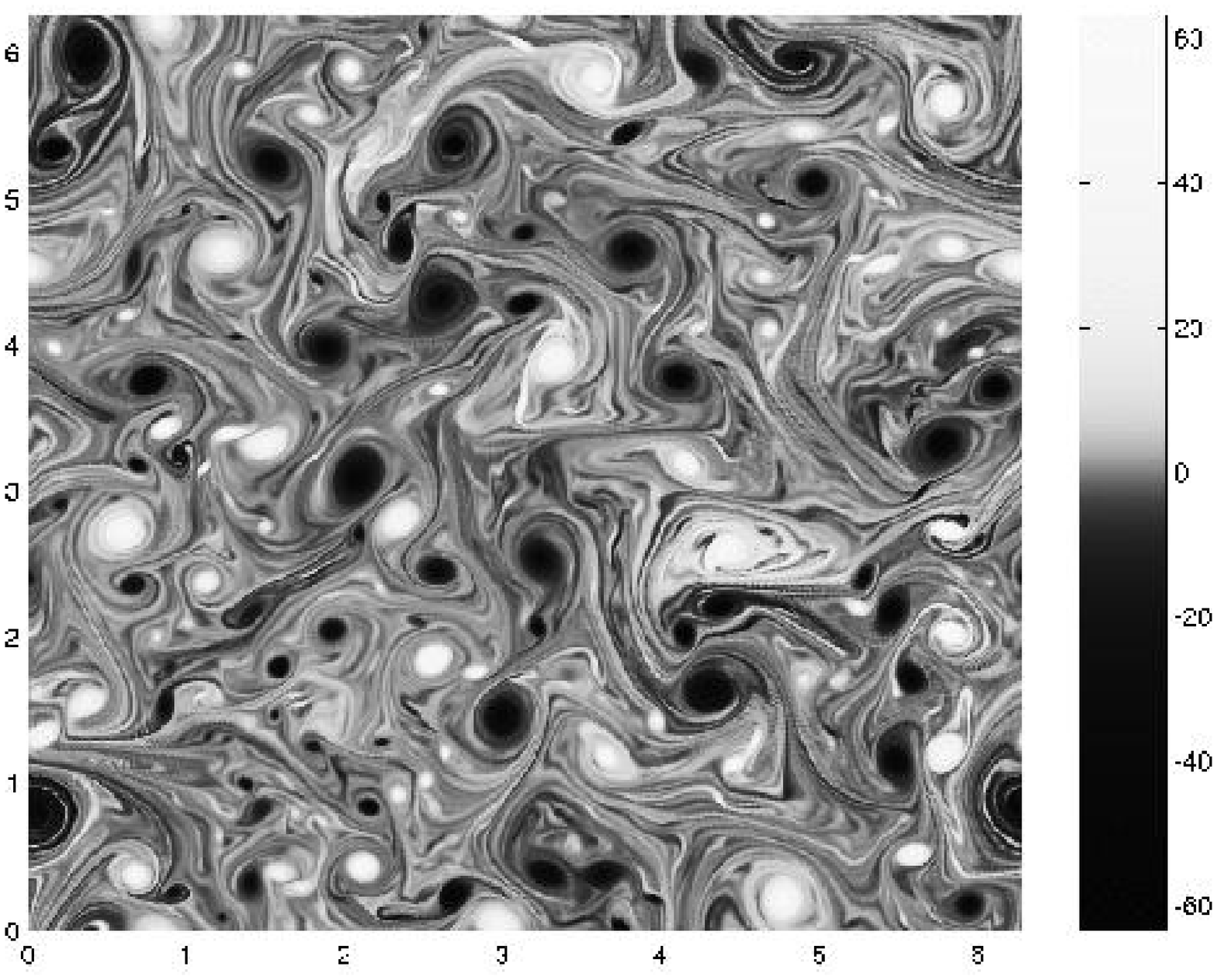}}
  \subfigure[R;~$\omega(x,y)$;~T=120]{\includegraphics[height=4cm]{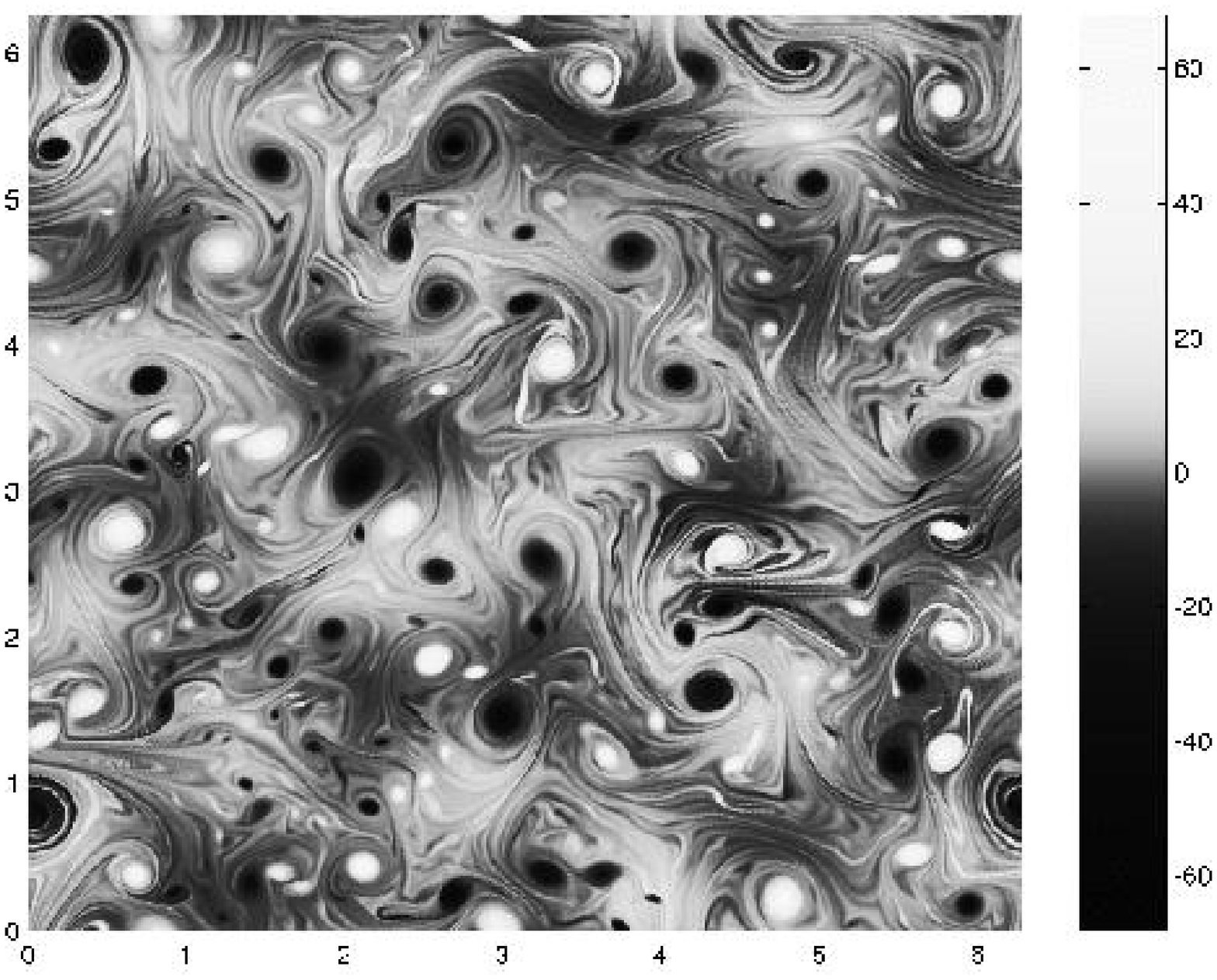}}
  \subfigure[N;~$\omega(x,y)$$|k|<5$ ]{\includegraphics[height=4cm]{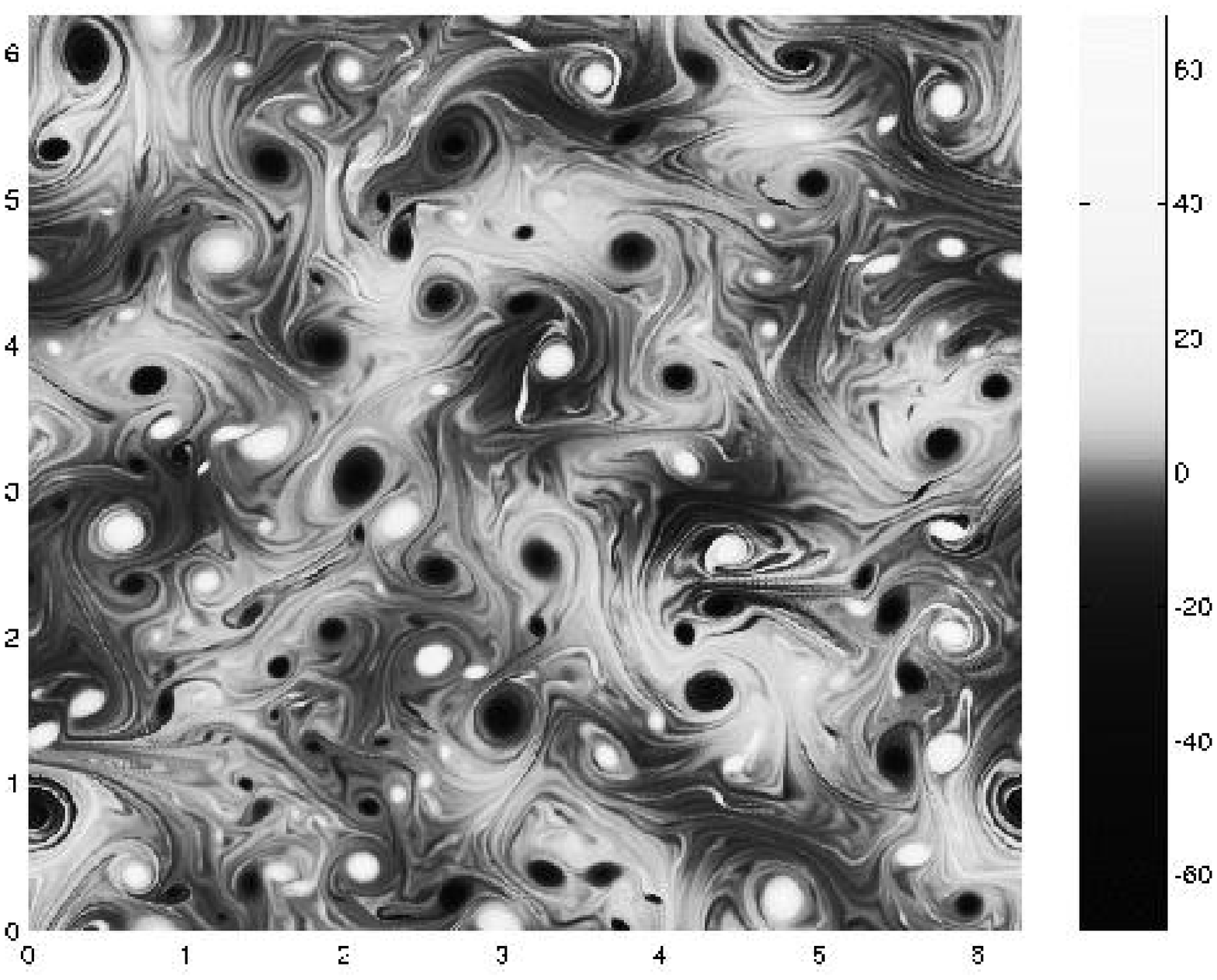}}
  \caption{Vorticity fields A,~R,~ and N at the transition time}
  \label{fig:omega_compare_1}
\end{figure}

\begin{figure}[h]
  \center
  \subfigure{\includegraphics[height=4cm]{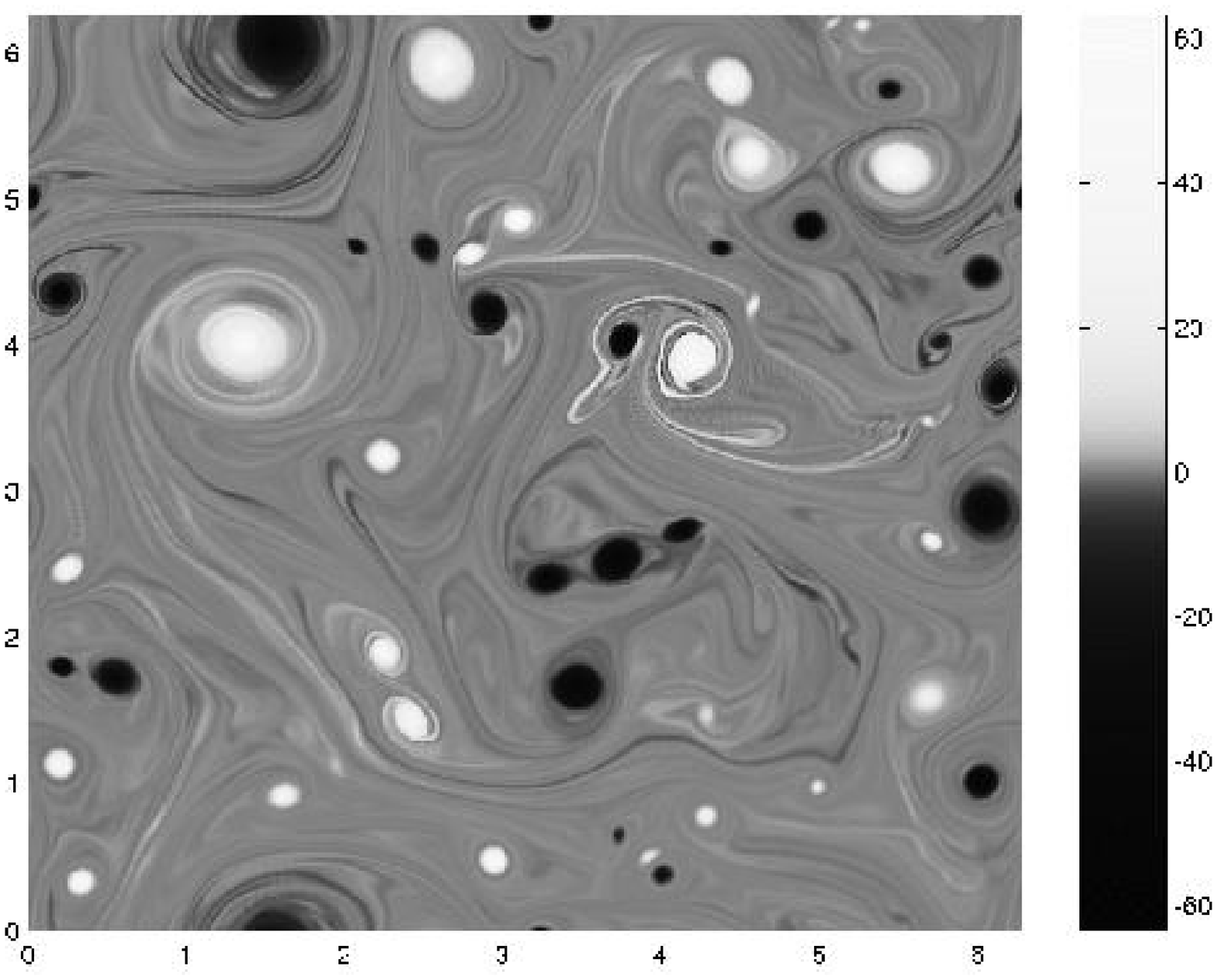}}
  \subfigure{\includegraphics[height=4cm]{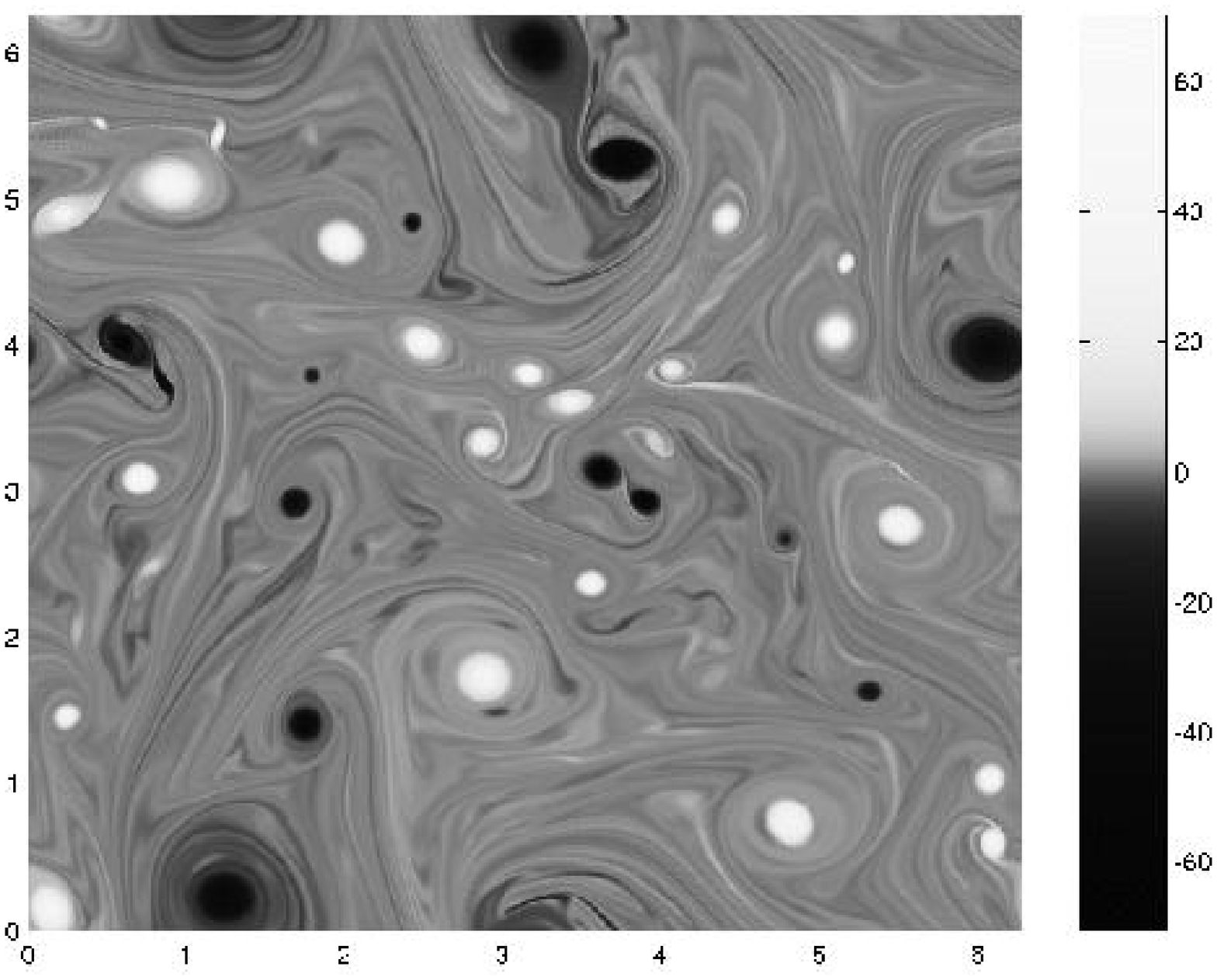}}
  \subfigure{\includegraphics[height=4cm]{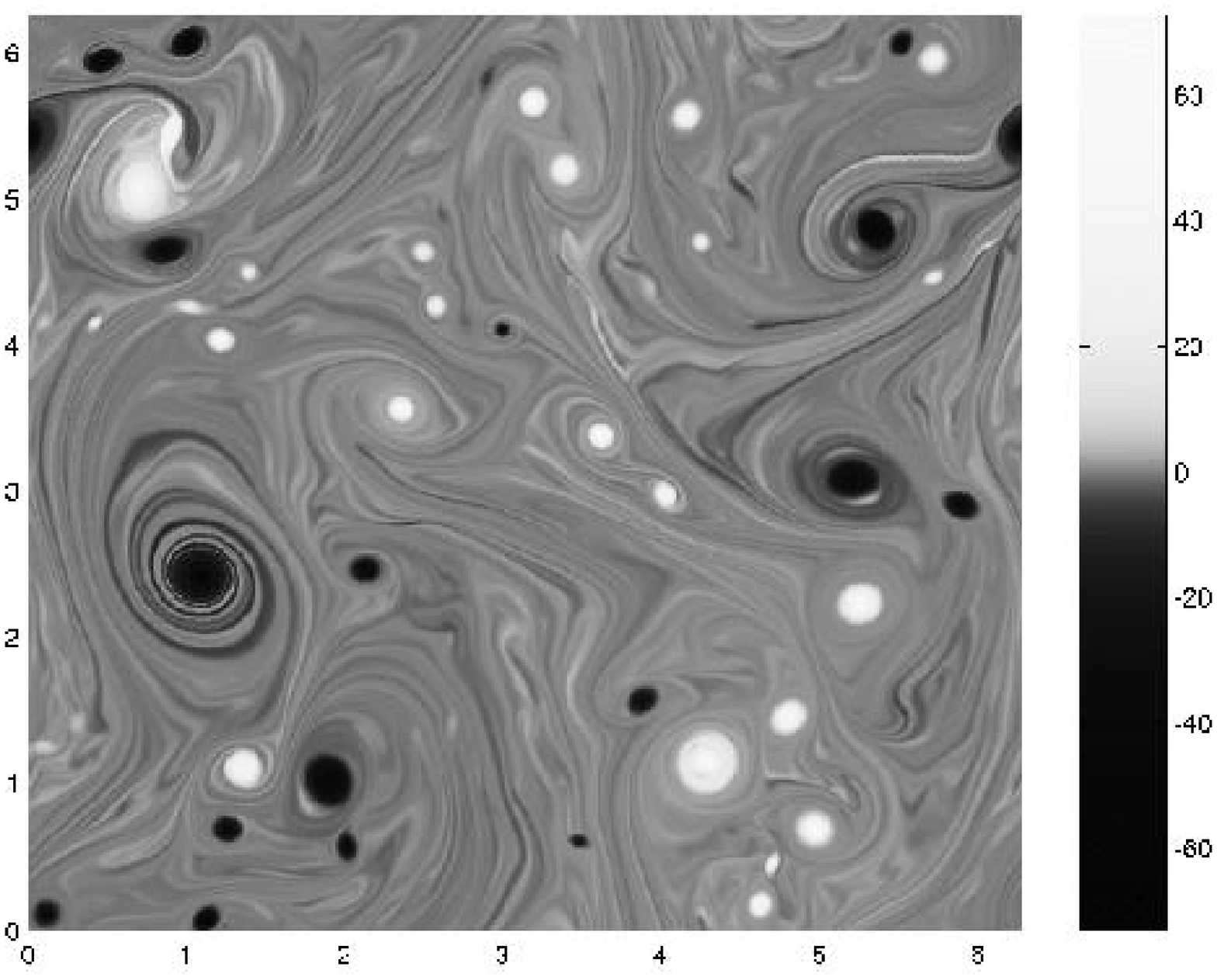}}
  \caption{Same as Fig. 2 but at time $T=480$}
  \label{fig:omega_compare_3}
\end{figure}

\begin{figure}[h]
  \center
  \subfigure{\includegraphics[height=4cm]{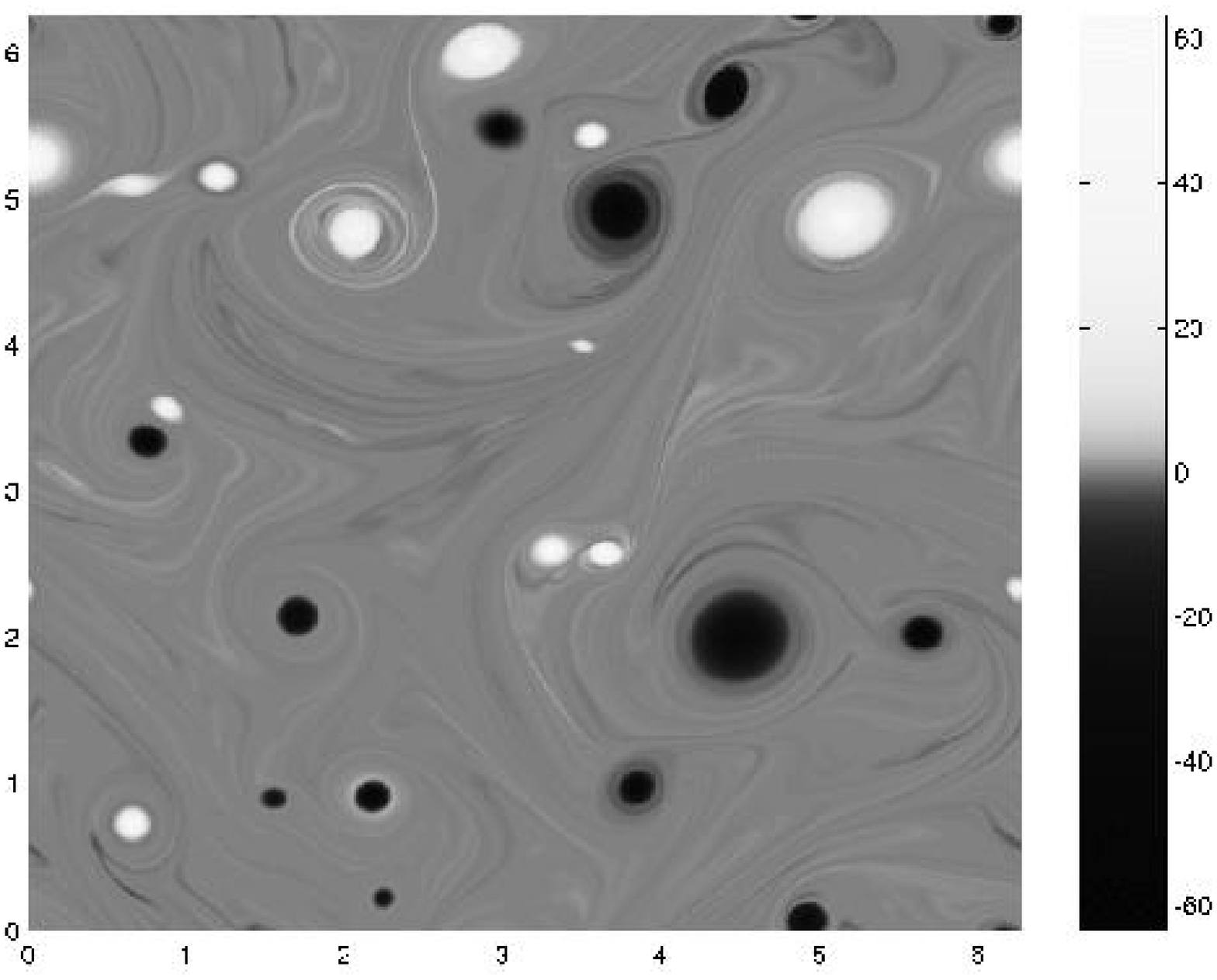}}
  \subfigure{\includegraphics[height=4cm]{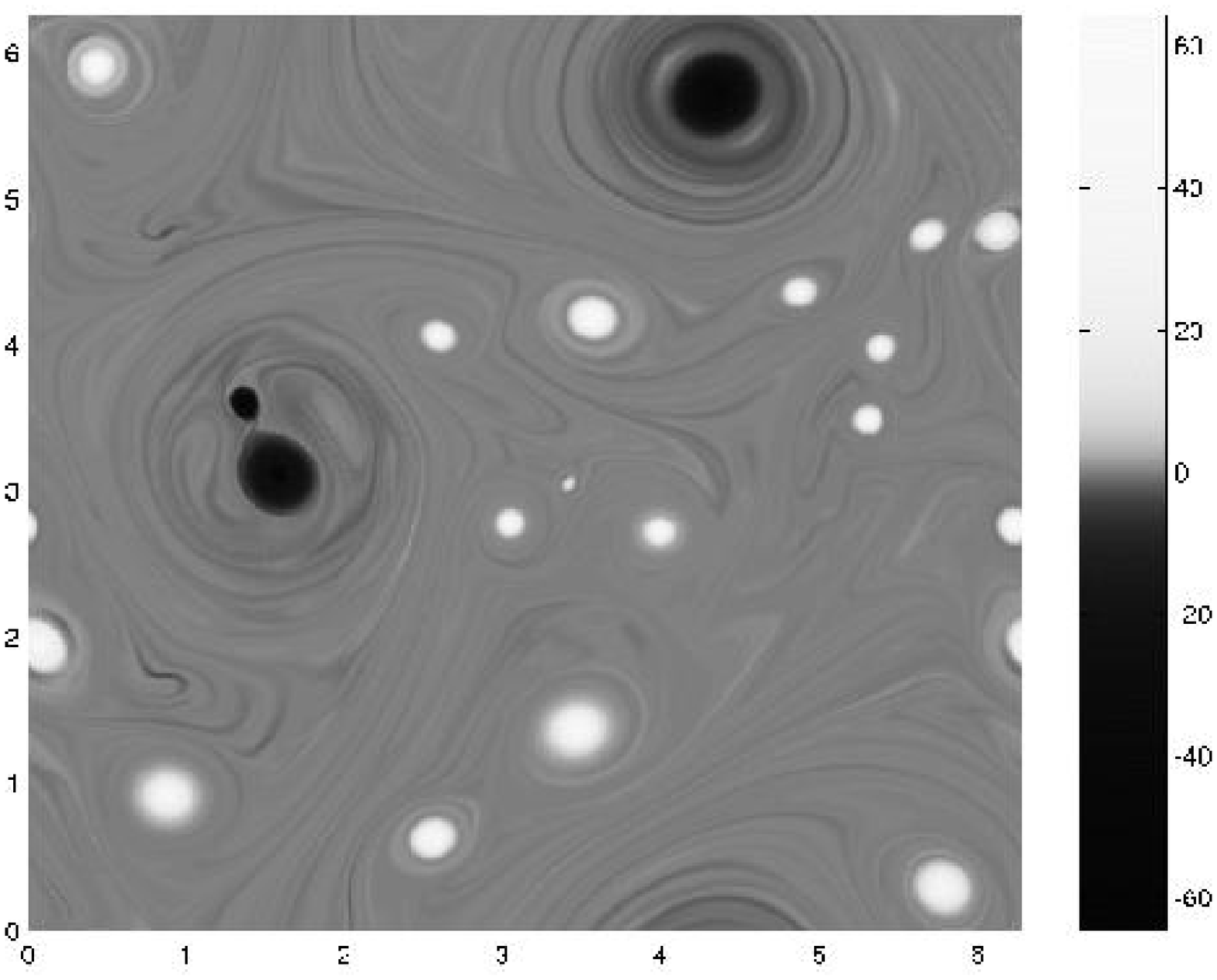}}
  \subfigure{\includegraphics[height=4cm]{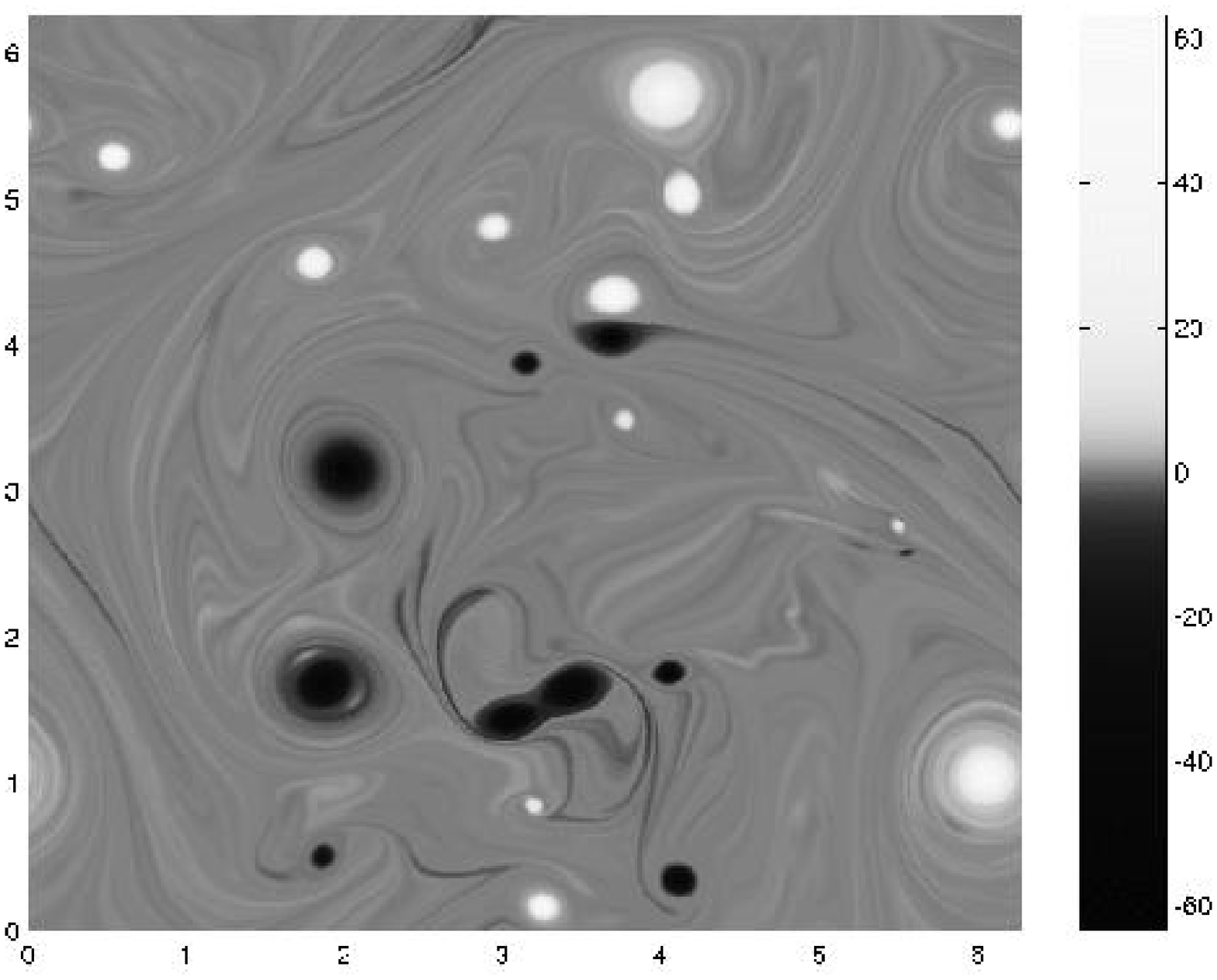}}
  \caption{same as Fig.2 but at $T=975$}
  \label{fig:omega_compare_4}
\end{figure}
  
\begin{figure}[h]
  \center
  \includegraphics[height=9cm]{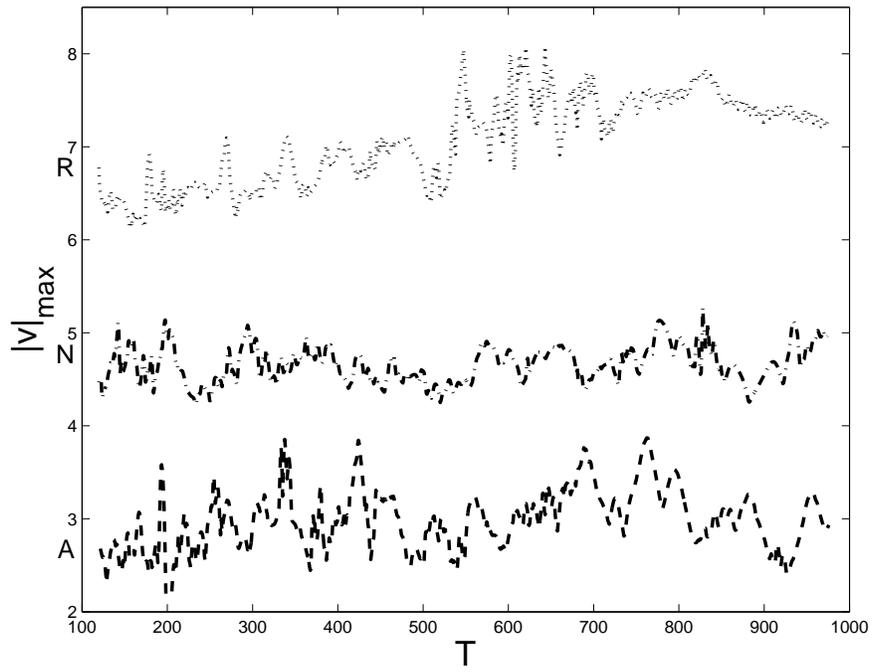}
  \caption{Time-evolution of maximum velocity for three runs: Curve $A$:~~$v_{max}(t)$; $R$:~~$v_{max}+4$  vs. t;  $N$:~~$v_{max}+2 $ vs t; }
  \label{fig:u_max_vs_t}
\end{figure}

\end{document}